# Nonlinearity-induced broadening of resonances in dynamically modulated couplers

A. Szameit,[1] Y. V. Kartashov,[2] M. Heinrich,[1] F. Dreisow,[1] R. Keil,[1] S. Nolte,[1] A. Tünnermann,[1] V. A. Vysloukh,[2] F. Lederer,[3] and L. Torner[2]

[1]Institute of Applied Physics, Friedrich Schiller University Jena, Max-Wien-Platz 1, 07743 Jena, Germany

[2]ICFO-Institut de Ciencies Fotoniques, and Universitat Politecnica de Catalunya, Mediterranean Technology Park, 08860 Castelldefels (Barcelona), Spain

[3]Institute for Condensed Matter Theory and Optics, Friedrich-Schiller-University Jena, Max-Wien-Platz 1, 07743 Jena, Germany

We report the observation of nonlinearity-induced broadening of resonances in dynamically modulated directional couplers. When the refractive index of the guiding channels in the coupler is harmonically modulated along the propagation direction and out-of-phase in two channels, coupling can be completely inhibited at resonant modulation frequencies. We observe that nonlinearity broadens such resonances and that localization can be achieved even in detuned systems at power levels well below those required in unmodulated couplers.

The precise control of wave packet evolution in systems with inhomogeneous refractive index landscapes is of major importance in optics [1,2]. When refractive index varies in both transverse and longitudinal directions a number of tools for controlling the propagation dynamics become available. Such structures can be used to manage diffraction properties in waveguide arrays [3,4], where diffraction-managed solitons can form [5,6]; it can be also used to drag solitons in the transverse plane [7,8], or to initiate soliton center oscillations and shape conversions [9,10]. Dynamic localization (DL) in waveguide arrays and coherent destruction of tunneling (CDT) in directional couplers, are possible either when guiding channels bend periodically in the propagation direction [11-16] or when the guiding channels remain straight, but their refractive index or widths oscillate periodically [17,18]. It turns out that DL and CDT are resonant effects, occurring only for specific modulation amplitude and frequency values. However, recently it was suggested that nonlinearity may cause a broadening of the resonance in directional couplers [19], so that the so-called nonlinear coherent destruction of tunneling (NCDT) yields a tunneling inhibition also in structures with a frequency slightly detuned from the resonance one and for powers well below the power required for the formation of a soliton in unmodulated system. In this Letter we experimentally study NCDT in a directional coupler where the refractive index is harmonically modulated along the propagation direction and out-of-phase in two channels. We observe that nonlinearity results in a broadening of resonances and that this broadening is proportional to the input beam power in a certain parameter range. We address here only the two-channel coupler but similar results may be encountered in more complicated systems such as waveguide arrays.

To gain insight into the phenomenon we analyze a theoretical model based on the nonlinear Schrödinger equation for the dimensionless field amplitude $q$, which governs the propagation of light beams along the $\xi$-axis of two-channel coupler under the assumption of cw radiation:

$$i\frac{\partial q}{\partial \xi} = -\frac{1}{2}\frac{\partial^2 q}{\partial \eta^2} - |q|^2 q - pR(\eta,\xi)q. \tag{1}$$

Here $\eta$ and $\xi$ are the transverse and longitudinal coordinates normalized to the characteristic transverse scale and diffraction length, respectively, while the parameter $p$ describes refractive index modulation depth. The refractive index distribution in the coupler is given by $R(\eta,\xi) = [1 + \mu\sin(\Omega\xi)]\exp[-(\eta + w_\mathrm{s}/2)^6/w_\eta^6] + [1 - \mu\sin(\Omega\xi)]\exp[-(\eta - w_\mathrm{s}/2)^6/w_\eta^6]$, where a super-Gaussian refractive index profile of each waveguide is fitted to the shape of the real laser-written waveguides in our sample [20], $\mu$ describes longitudinal modulation depth, and $\Omega$ is the longitudinal modulation frequency. The parameter $w_\eta$ characterizes the waveguide width, while $w_\mathrm{s}$ stands for the separation between the coupler channels. We set $w_\eta = 0.3$ that is equivalent to 3 $\mu$m width of the waveguides, and $w_\mathrm{s} = 3.2$ that is equivalent to the separation of channels of 32 $\mu$m. We also set $p = 2.78$ that corresponds to a real modulation depth of refractive index $\sim 3.1\times10^{-4}$. In all simulations we used the input beams $q|_{\xi=0} = Aw(\eta)$, where $w(\eta)$ describes the shape of the fundamental linear mode of a single waveguide, and $A$ is the input amplitude.

In unmodulated linear couplers ($\mu = 0$) light periodically switches between both channels if only one of them is excited at $\xi = 0$. We define the linear beating period $T_b$ as the distance at which light returns to the input waveguide after one complete switching cycle (accordingly, the beating frequency is given by $\Omega_b = 2\pi/T_b$). For our set of parameters one has $T_b = 100$. A longitudinal out-of-phase modulation of the refractive index in the coupler channels results in the inhibition of coupling that takes place only for properly selected values of the modulation frequency $\Omega$ [15-17]. To characterize localization, we use a distance-averaged power fraction trapped in the excited channel $U_m = L^{-1} \int_0^L d\xi \int_{-\infty}^0 |q(\eta,\xi)|^2 d\eta \Big/ \int_{-\infty}^0 |q(\eta,0)|^2 d\eta$, where $L$ is the final distance. The dependence $U_m(\Omega)$ is characterized by multiple parametric resonances. Here we only consider the principal resonance with the largest frequency $\Omega_r$ (Fig. 1). The resonance frequency grows with increasing longitudinal modulation depth $\mu$ [Fig. 1(a)], and for moderate $\mu$ values $\Omega_r \sim \mu$. In the limit $\mu \to 0$ one has $\Omega_r \to \Omega_b/2$. The half-width of the principal resonance $\delta\Omega/\Omega_b$ defined at the level $U_m = 0.7$ is a monotonically decreasing function of the averaging distance $L$ [Fig. 1(b)]. To understand the impact of nonlinearity on resonance curve it is desirable to perform averaging over sufficiently long distance $L$. However, the value $\delta\Omega/\Omega_b$ at $L \to \infty$ is determined mostly by the amplitude $A$ of the input beam when nonlinearity is taken into account since in this limit a vanishing resonance width would only occur for a vanishing amplitude of the propagating beam. In contrast, in our calculations the resonance width is always finite due to the nonzero power used in the initial conditions. In Fig. 1(c) it is shown how a variation of $\mu$ results in a shift of the principal resonance. Around the resonance the dependence $U_m(\Omega)$ is close to a $\text{sinc}[\alpha L(\Omega - \Omega_r)]$ function superimposed on a constant pedestal (this is a direct consequence of averaging of an oscillating function over a large distance).

Figure 2 illustrates how the resonance is affected by the nonlinearity. The important result is that the resonance broadens with increase of input amplitude $A$, which is shown in Fig. 2(a). The resonance width increases almost linearly with input power $\sim A^2$ corresponding to the findings in Ref. [20]. As $A \to 0$ the width of resonance approaches a minimal value (corresponding to the resonance width in the linear medium) that decreases with increasing averaging distance $L$ [see, e.g., Fig. 1(b)]. The resonance broadening is illustrated in Fig. 2(c) where the curves 1 and 2 correspond to small and moderate $A$ values, respectively. The broadening of resonance can be considerable even at moderate amplitudes, which are well below the critical amplitude $A_{\rm cr} \sim 0.42$ required for coupling suppression in unmodulated system.

To confirm these trends experimentally we fabricated a sequence of directional couplers using the femtosecond laser writing technique [4] with laser pulses at $\lambda = 800$ nm, a temporal width of 140 fs and a repetition rate of 100 kHz. The sample material of choice was fused silica of highest quality, the focusing was achieved by 20x objective (NA $= 0.45$). For the experimental analysis of the predicted resonance broadening, we launched light at $\lambda = 800$ nm in one waveguide using a 2.5x objective (NA $= 0.05$). The end facet of the sample was imaged on a CCD camera, using a 10x objective (NA $= 0.25$). The induced index change $\delta n$ depends on the writing speed $v$ approximately as $\delta n \approx \alpha \exp(-v/\beta) + \gamma$, where the constants $\alpha, \beta, \gamma$ are fitted numerically to the experimental results [20]. A sinusoidal index change along the individual guides is then achieved by varying writing speed as $v = \beta \ln(\alpha/[p + \mu \sin(\Omega z) - \gamma])$. The sample length was 105 mm and waveguide spacing was 32 $\mu$m. The beating period is $\sim 115$ mm that corresponds to $T_{\rm b} = 100$. The guides exhibit an average refractive index of $\sim 3.1 \times 10^{-4}$ and a modulation amplitude of $\approx 2 \times 10^{-5}$. For these parameters, the modulation frequency corresponding to principal resonance amounts to $\Omega_{\rm r} \approx 0.180$ mm$^{-1}$ (i.e., $\Omega_{\rm r} \approx 3.27 \Omega_{\rm b}$). To analyze localization in the vicinity of $\Omega_r$, we fabricated couplers with modulation frequencies $\Omega = 0.180, 0.176, 0.173, 0.170, 0.167, 0.165,$ and $0.162$ mm$^{-1}$.

The width of the resonance curve $\delta\Omega/\Omega_b$ was defined using those powers, where 70% of input power still remains in the excited channel. In the resonant coupler a minimal power of 50 kW was injected to achieve localization. The powers for the detuned couplers defined with the above mentioned criterion are:

| $\Omega$ [mm$^{-1}$] | $P$ [kW] |
| --- | --- |
| 0.180 | 50 |
| 0.176 | 140 |
| 0.173 | 250 |
| 0.170 | 300 |
| 0.167 | 340 |
| 0.165 | 370 |
| 0.162 | 400 |

as shown in Fig. 2(b). The circles represent measured points, including the error of about 15 kW. For very small detuning the dependency is nonlinear, as confirmed by the continuous model [see Fig. 2(a)]. However, $P(\delta\Omega)$ dependence becomes almost linear as detuning increases, which is consistent with results of Ref. [20] obtained for discrete model. Representative output patterns for couplers with different modulation frequencies are shown in Fig. 4. Note that in all cases the left waveguide was excited. In the left column [Fig. 4(a)] the coupler at the resonance frequency is shown. For all applied input powers, the light remains localized. In contrast, in detuned coupler with $\Omega = 0.173$ mm$^{-1}$ [Fig. 4(b)] the light at 50 kW couples strongly into second guide (first row). At 250 kW input power, about 70 % of the injected light remains in the excited channel (second row), defining the width of the resonance curve. For higher peak powers (e.g., 340 kW in third row), light almost completely localizes in the excited waveguide. When frequency is detuned from resonance even farther, i.e. when $\Omega = 0.167$ mm$^{-1}$, at 50 kW a large fraction of the light couples into the second waveguide [see first row of Fig. 4(c)]. Even at 250 kW the localization is just slightly larger (second row). More than 70 % of the light remains in the excited guide only above an input power of 340 kW (third row), which then again defines the width of the resonance curve.

    In conclusion, we observed resonance-broadening in longitudinally modulated couplers induced by nonlinearity. Also, localization was observed in couplers with a modulation frequency considerably detuned from resonance at powers well below the threshold required in unmodulated system. We confirmed that the dependence of power on resonance width is approximately linear.

# References with titles

# References without titles

# Figure captions

Figure 1.  (a) Resonance frequency versus $\mu$. (b) The half-width of resonance curve defined at the level $U_\mathrm{m} = 0.7$ versus coupler length at $\mu = 0.19$ and $A = 0.1$. (c) $U_\mathrm{m}$ versus $\Omega/\Omega_\mathrm{b}$ at $A = 0.1$, $L = 8T_\mathrm{b}$ for modulation depth $\mu = 0.15$ (curve 1) and $\mu = 0.19$ (curve 2).

Figure 2.  (a) Theoretically calculated $A^2$ value versus half-width of resonance curve defined at the level $U_\mathrm{m} = 0.7$ at $\mu = 0.19$ and $L = 8T_\mathrm{b}$. (b) Experimentally obtained dependence of input peak power required for $70\%$ localization in the launching channel versus normalized detuning from resonance frequency. (c) $U_\mathrm{m}$ versus $\Omega/\Omega_\mathrm{b}$ at $\mu = 0.19$, $L = 8T_\mathrm{b}$ for input amplitudes $A = 0.03$ (curve 1) and $A = 0.15$ (curve 2).

Figure 3.  Propagation dynamics in (a) unmodulated coupler at $\mu = 0.00$, $A = 0.01$, and modulated couplers at (b) $\mu = 0.19$, $A = 0.01$, (c) $\mu = 0.19$, $A = 0.12$, (d) $\mu = 0.19$, $A = 0.24$. Propagation distance is $L = 8T_\mathrm{b}$, while modulation frequency in (b)-(d) is $\Omega = 3.10\Omega_\mathrm{b}$. In all cases left waveguide is excited at $\xi = 0$.

Figure 4.  Output intensity distributions for an excitation of left channel of modulated coupler when (a) $\Omega = 3.27\Omega_\mathrm{b}$, (b) $\Omega = 3.15\Omega_\mathrm{b}$, and (c) $\Omega = 3.04\Omega_\mathrm{b}$. The input peak power is $50 \mathrm{~kW}$ in first row, $250 \mathrm{~kW}$ in second row, $340 \mathrm{~kW}$ in third row.

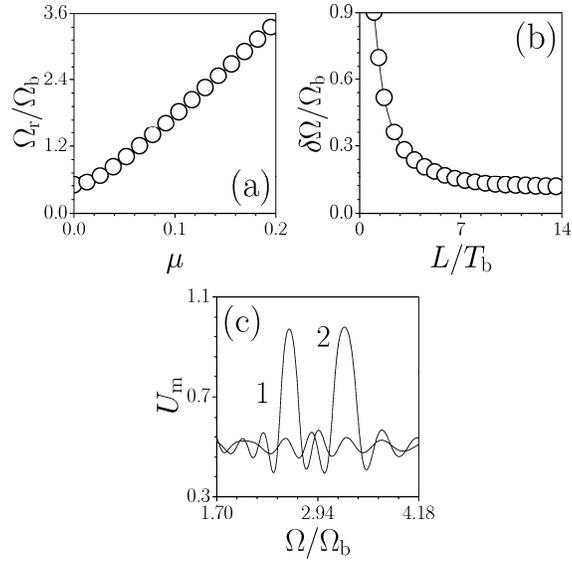

Figure 1.   (a) Resonance frequency versus $\mu$. (b) The half-width of resonance curve defined at the level $U_\mathrm{m} = 0.7$ versus coupler length at $\mu = 0.19$ and $A = 0.1$. (c) $U_\mathrm{m}$ versus $\Omega/\Omega_\mathrm{b}$ at $A = 0.1$, $L = 8T_\mathrm{b}$ for modulation depth $\mu = 0.15$ (curve 1) and $\mu = 0.19$ (curve 2).

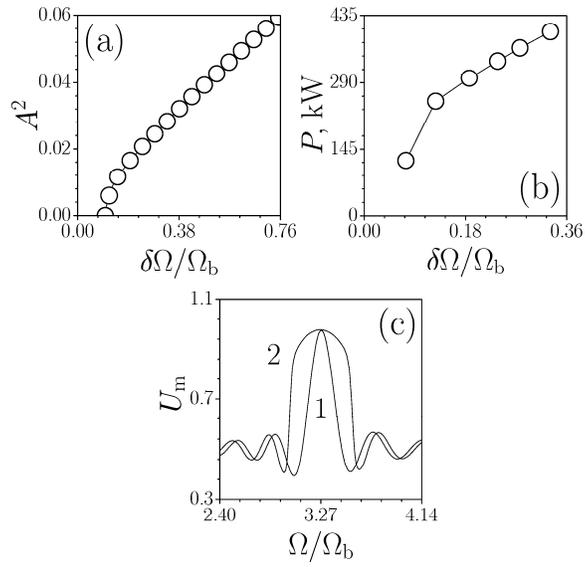

Figure 2. (a) Theoretically calculated $A^2$ value versus half-width of resonance curve defined at the level $U_{\mathrm{m}} = 0.7$ at $\mu = 0.19$ and $L = 8T_{\mathrm{b}}$. (b) Experimentally obtained dependence of input peak power required for 70% localization in the launching channel versus normalized detuning from resonance frequency. (c) $U_{\mathrm{m}}$ versus $\Omega/\Omega_{\mathrm{b}}$ at $\mu = 0.19$, $L = 8T_{\mathrm{b}}$ for input amplitudes $A = 0.03$ (curve 1) and $A = 0.15$ (curve 2).

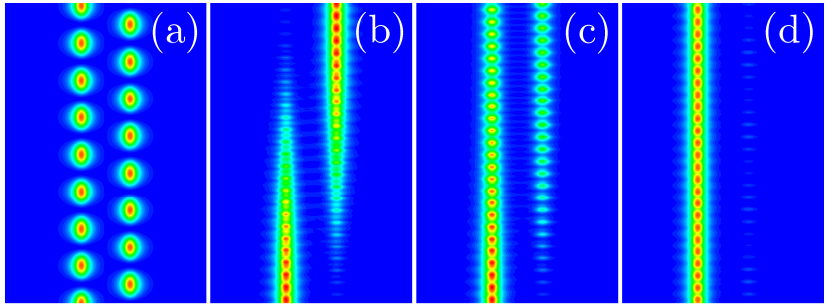

Figure 3.  Propagation dynamics in (a) unmodulated coupler at $\mu = 0.00$, $A = 0.01$, and modulated couplers at (b) $\mu = 0.19$, $A = 0.01$, (c) $\mu = 0.19$, $A = 0.12$, (d) $\mu = 0.19$, $A = 0.24$. Propagation distance is $L = 8T_{\rm b}$, while modulation frequency in (b)-(d) is $\Omega = 3.10\Omega_{\rm b}$. In all cases left waveguide is excited at $\xi = 0$.

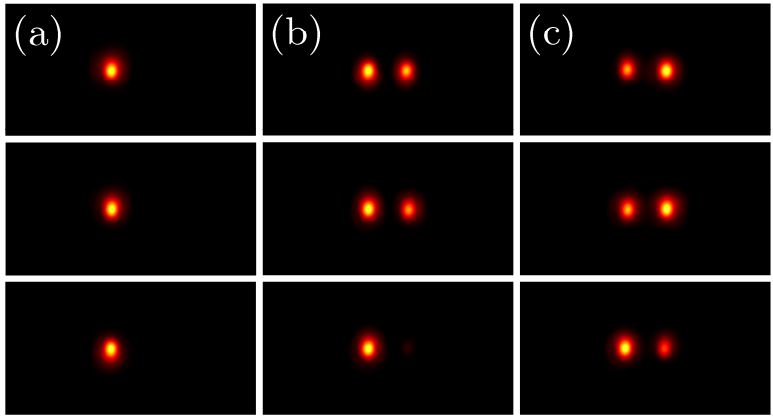

Figure 4.  Output intensity distributions for an excitation of left channel of modulated coupler when (a) $\Omega = 3.27\Omega_{\text{b}}$, (b) $\Omega = 3.15\Omega_{\text{b}}$, and (c) $\Omega = 3.04\Omega_{\text{b}}$. The input peak power is 50 kW in first row, 250 kW in second row, 340 kW in third row.